\documentclass[10pt,conference]{IEEEtran}
\usepackage{cite}
\usepackage[dvips]{color}
\usepackage{tabu}
\usepackage{comment}
\usepackage{todonotes}
\usepackage{epsf}
\usepackage{epsfig}
\usepackage{times}
\usepackage{epsfig}
\usepackage{graphicx}

\usepackage{mathtools}
\usepackage{mathrsfs}
\usepackage{amssymb,amsmath}
\usepackage{amsmath}
\usepackage{amsfonts}

\usepackage{algorithmicx}  
\usepackage[algo2e]{algorithm2e} 
\usepackage{cleveref}
\usepackage{dsfont}
\usepackage{lettrine} 
\usepackage{epsfig,algorithm,algpseudocode,amsthm,url}
\usepackage{caption}
\usepackage{subcaption}
\usepackage[justification=raggedright]{caption}
\usepackage{subcaption}
\usepackage{bbm}
\usepackage[utf8]{inputenc}
\usepackage{verbatim}
\usepackage{setspace}	
\usepackage[T1]{fontenc}
\usepackage{textcomp}
\usepackage{tabularx}
\usepackage{xpatch}
\DeclareMathOperator*{\argmax}{argmax}

\usepackage{stackengine}
\usepackage[nocomma]{optidef}
\usepackage{commath}

\usepackage[left=1.62cm,right=1.62cm,top=1.85cm,bottom=2.6cm]{geometry}

\IEEEoverridecommandlockouts 

  \renewcommand{\baselinestretch}{0.95}

\begin{document}
\title{\Huge Meta Reinforcement Learning Approach for Adaptive Resource Optimization in O-RAN \thanks{This material is based upon work supported by the National Science Foundation under Grant Numbers  CNS-2202972, CNS- 2318726, and CNS-2232048.} 
}\vspace{-0.3cm}

\author{
	\IEEEauthorblockN{
	Fatemeh Lotfi, Fatemeh Afghah}

	\IEEEauthorblockA{Holcombe Department of Electrical and Computer Engineering, Clemson University, Clemson, SC, USA \\
Emails: flotfi@clemson.edu,   fafghah@clemson.edu} }\vspace{-0.2cm}

\maketitle\vspace{-0.5cm}
\begin{abstract} 
As wireless networks grow to support more complex applications, the Open Radio Access Network (O-RAN) architecture, with its smart RAN Intelligent Controller (RIC) modules, becomes a crucial solution for real-time network data collection, analysis, and dynamic management of network resources including radio resource blocks and downlink power allocation. 
Utilizing artificial intelligence (AI) and machine learning (ML), O-RAN addresses the variable demands of modern networks with unprecedented efficiency and adaptability. Despite progress in using ML-based strategies for network optimization, challenges remain, particularly in the dynamic allocation of resources in unpredictable environments. This paper proposes a novel Meta Deep Reinforcement Learning (Meta-DRL) strategy, inspired by Model-Agnostic Meta-Learning (MAML), to advance resource block and downlink power allocation in O-RAN. 
Our approach leverages O-RAN’s disaggregated architecture with virtual distributed units (DUs) and meta-DRL strategies, enabling adaptive and localized decision-making that significantly enhances network efficiency. By integrating meta-learning, our system quickly adapts to new network conditions, 
optimizing resource allocation in real-time. This results in a \(19.8\%\) improvement in network management performance over traditional methods, advancing the capabilities of next-generation wireless networks.

\end{abstract}\vspace{-0.1cm}
\section{Introduction} 

The advancement of wireless networks to support diverse and demanding applications is greatly enhanced by the Open Radio Access Network (O-RAN) architecture, particularly its RAN Intelligent Controller (RIC) modules \cite{polese2022understanding, 3gppRe18}. These modules boost network functionality through intelligent resource management and sophisticated control techniques, essential for delivering advanced services by enabling real-time data collection and analysis \cite{polese2022understanding}. Moreover, integrating artificial intelligence (AI) and machine learning (ML) within these modules facilitates dynamic resource allocation, enhancing operational efficiency and adaptability to rapidly changing conditions. Central to this innovation, RIC modules employ open and standardized interfaces for both real-time and non-real-time control, making the network more intelligent, fully virtualized, and interoperable \cite{d2022dapps}. 

ML-based strategies, particularly for adaptive network configurations, are crucial in the dynamic realm of wireless networks. The RIC's ability to utilize key performance indicators (KPIs) and perform real-time service analysis enables the network to adjust to fluctuating demands dynamically. Despite extensive research into ML-based power and resource allocation, significant challenges remain in managing the complexities of real-time resource management under unpredictable conditions ~\cite{ji2023meta,raftopoulos2024drl,10071958,cheng2022reinforcement,thaliath2022predictive}. These challenges are exacerbated in the dynamic and complex landscape of O-RAN, where the network must rapidly adapt by integrating virtualized distributed units (DUs) and deploying xApps based on real-time analysis. Scheduling presents a particularly acute challenge in this context, especially in densely populated or hotspot areas where demand can surge unexpectedly and where corner cases, scenarios beyond standard operating parameters, commonly occur. This unpredictability necessitates highly adaptive scheduling systems capable of responding swiftly to changing conditions. To address these critical needs, we have chosen the enhanced mobile broadband (eMBB) service slice, known for its high data rate requirements, as a representative use case in densely populated areas. This selection highlights the imperative for advanced scheduling solutions that can adeptly handle and respond to users' unpredictable and varied demands.

Advanced scheduling strategies are essential to address these intricate challenges, including managing corner cases and achieving rapid convergence. We utilize meta-learning in a scheme that significantly enhances the efficiency and adaptability of Deep Reinforcement Learning (DRL) approaches, proving particularly advantageous in few-shot learning scenarios where deriving meaningful insights from small data samples is crucial~\cite{beck2023survey,yuan2021meta,erdol2022federated,ji2023meta}. Unlike federated learning (FL), which focuses on decentralized model training without exchanging local data, meta-learning optimizes the learning algorithm to enhance its ability to generalize from past experiences to new situations. Meta-learning stands out by leveraging this prior knowledge to accelerate the learning process, distilling reusable skills to facilitate rapid adaptation to new tasks. This is especially beneficial in the dynamic landscapes of wireless networks, where conditions and service demands frequently shift and are often application-dependent. Building on this foundation, we develop a novel meta-learning-based joint resource block and downlink power allocation solution inspired by Model-Agnostic Meta-Learning (MAML) in advanced wireless O-RAN network architectures \cite{Owfi}. Implemented within advanced wireless O-RAN architectures, our strategy mitigates system performance disruptions. It ensures optimal service delivery to user equipment (UEs), establishing a new network optimization standard in highly dynamic environments.

Our architecture capitalizes on the disaggregated structure of O-RAN by strategically placing distributed DRL agents at DU locations for targeted learning tasks. This configuration exploits the distributed nature of DU modules, each augmented with individual xApps, to achieve remarkable scalability, adaptability, and localized decision-making—crucial for navigating the complex and dynamic landscapes of O-RAN environments. Our design significantly enhances the network’s responsiveness and operational efficiency as it scales by minimizing overhead and latency through decisions made close to data sources. Furthermore, our emphasis on localized processing significantly improves out-of-order messaging capabilities, essential for maintaining reliability and performance in sophisticated network scenarios. Additionally, our proposed meta-DRL approach allows the system to adapt to evolving conditions swiftly, optimizing resource allocation in near real-time and enhancing reliability. By integrating a meta-learning strategy, our approach enables distributed DRL agents deployed across the O-RAN architecture to manage resource allocation and network operations efficiently. They rapidly adapt to new tasks with minimal training, optimizing decision-making processes and marking a significant advancement in O-RAN architecture, paving the way for more efficient and resilient next-generation wireless networks. In a dynamically configured O-RAN environment featuring distributed DUs and mobile users, our simulation assesses the impact of our meta-learning strategy. The simulation results reveal a remarkable $19.8\%$ improvement in the final return value for intelligent network management compared to baseline methods. \emph{To the best of our knowledge, this is the first work that develops a MAML-inspired RL technique for joint resource block and downlink power allocation in O-RANs}.

The structure of this paper is as follows: Section \ref{related} discusses the previous works on utilizing meta-learning approaches in the context of DRL for resource allocation issues. Section \ref{sysmodl} details the system model and problem formulation for O-RAN resource and power allocation. Section \ref{MDRL} describes our MAML-inspired Meta-DRL solution. Simulation outcomes are shared in Section \ref{sim}, with final conclusions in Section \ref{conclusion}.\vspace{-0.2cm}

\begin{figure}[t!]
  \centering
    \includegraphics[width=0.76\columnwidth]{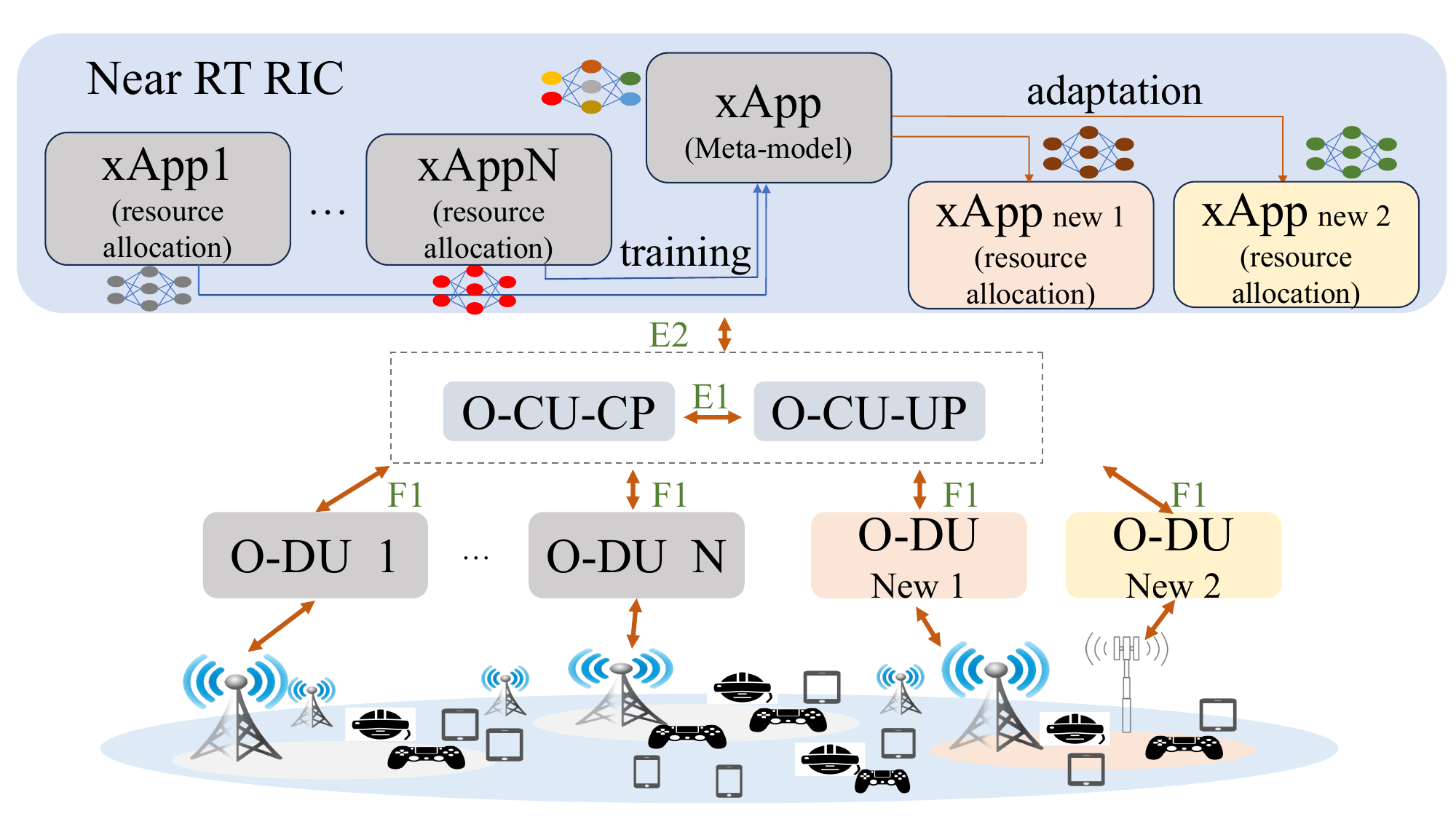}
    \caption{\small System model topology for MAML-inspired DRL resource block and downlink power allocation in O-RAN architecture.  
    }
    \label{sys_graph}
\end{figure}
\section{Related Works}\label{related}
The evolution of wireless networks necessitates advancements in O-RAN architecture, particularly by adopting intelligent frameworks such as meta-learning and DRL, to enhance network efficiency and adaptability. The study by Erdol et al.~\cite{erdol2022federated} demonstrates a meta-federated RL-based traffic steering algorithm within the non-real-time RIC (non-RT RIC) for Radio Access Technology (RAT) allocation aimed at fulfilling users' Quality of Service (QoS) requirements. This algorithm employs Deep Q-Networks (DQN) to dynamically steer traffic between different RATs, thus optimizing network performance and user experience. While federated RL offers potential for distributed learning, meta-RL stands out in O-RAN scenarios for its quick adaptability and seamless architectural fit, promising more effective scalability and implementation. Furthermore, Ji et al.~\cite{ji2023meta} discuss distributed resource allocation strategies to maximize energy efficiency and maintain QoS for UEs through a meta-federated RL approach. This method optimizes both channel assignment and transmission power concurrently. In parallel, Yuan et al.~\cite{yuan2021meta} explore the application of meta-federated RL to solve resource and power allocation challenges within vehicle-to-everything (V2X) communication scenarios. However, the approaches by Ji et al. and Yuan et al., which utilize power quantization for allocation, are noted to introduce a certain error level. 

Additionally, research presented in~\cite{10071958} outlines a sequential method for resource optimization, which, despite its potential, may yield suboptimal solutions due to a failure to account for interdependencies between parameters. This method's iterative nature, focusing on fixing and optimizing parameters in isolation, could lead to slower convergence and inferior outcomes. Kalntis et al.~\cite{kalntis2023adaptive} introduce an adaptive meta-learning-based online learning framework for resource allocation within O-RAN's virtualized Base Stations (vBS), showcasing the framework's adaptability to network changes. Nevertheless, the simplicity of their policy selection and the static nature of their power allocation suggest that further enhancements are necessary to fully exploit O-RAN's dynamic capabilities. Lastly, Raftopoulos et al.~\cite{raftopoulos2024drl} delve into managing fluctuating Service Level Agreements (SLAs), highlighting the challenge of balancing latency requirements against other network demands as an area ripe for further exploration. Together, these studies point towards developing more autonomous, efficient, and intelligent O-RAN architectures, underscoring the importance of comprehensive strategies that integrate diverse technologies to optimize a wide range of network performance metrics. \vspace{-0.2cm}

\section{System model}\label{sysmodl}


In our envisioned O-RAN architecture for a wireless communication network, we integrate multiple Central Units (CUs), including the Control Plane (CU-CP) for network orchestration and the User Plane (CU-UP) for data processing, along with DUs and Radio Units (RUs) covering macro and small cells. All components are coordinated by the RIC to dynamically allocate resource blocks and downlink power, as illustrated in Fig.~\ref{sys_graph}. 
\vspace{-0.2cm}

\subsection{Achievable Data Rate}
We consider a downlink orthogonal frequency-division multiple access (OFDMA) system where a single CU module oversees multiple DU/RU paired units. Within this framework, each DU/RU unit is pre-configured and equipped with its own $K_\rho$ dedicated resource blocks (RBs) to serve $N_\rho$ assigned UEs. These units are integrated into the network infrastructure and connect dynamically as needed based on real-time demand or operational requirements.  
To address the challenge of fair service delivery, especially in densely populated areas with high demand for eMBB services, we frame our system goal as a $\max$-$\min$ problem.  
This approach aims to maximize the minimum data rate among all UEs, ensuring that no single user suffers from extremely poor service due to the skewed allocation of resources. Such an approach is crucial in maintaining service fairness, as it guarantees that the network's resources are distributed in a manner that lifts the lowest-performing connections to an acceptable standard. Each UE's data rate, significantly influenced by the allocated RBs and transmission power, becomes a focal point of our optimization efforts. Given $K_\rho$ RBs, and the maximum ($P_{max}$) and minimum ($P_{min}$) transmission powers, we navigate through the constraints of available RBs and downlink transmission power to ensure efficient and equitable resource distribution.

The achievable data rate of each UE $u$ from the assigned RU and DU $\rho$, by considering RU-UE channels as Rayleigh fading and assuming the proposed system model works in a discrete time frame as $t\in[1,2,..,T]$ can be expressed as:\vspace{-0.2cm}
\begin{align}\label{urate} 
    c_{u,\rho}(t) = \sum_{k=1}^{K_{\rho}} B e_{u,k}  \log \Big(1+\frac{p_{k,\rho}(t) d_{u,\rho}(t)^{-\eta} |h_{u,k}(t)|^2}{ I_{u,k}(t)+ \sigma^2}\Big),
\end{align}
where $e_{u,k} \in \{0, 1\}$ is a binary variable indicating RB allocation for user $u$ in RB $k$, and $B$ represents RB bandwidth. 
$p_{k,\rho}$ denotes transmission power allocated to the RB $k$ of RU $\rho$, and $d_{u,\rho}(t)$ represents the distance between RU $\rho$ and UE $u$ in each time frame which is time-varying variable because of the UEs mobility. Moreover, $\eta$ shows the path loss exponent, and $|h_{u,k}(t)|^2$ indicates the time-varying Rayleigh fading channel gain each time. In equation \eqref{urate}, the term $I_{u,k}(t)=\sum_{\rho' \neq \rho} \sum_{u'\neq u} e_{u',k} p_{k,\rho'}(t) d_{u',\rho'}(t)^{-\eta} |h_{u',k}(t)|^2$, represents downlink interference from neighboring RUs on RB $k$, 
while $\sigma^2$ is indicative of the variance associated with the additive white Gaussian noise (AWGN). \vspace{-0.2cm}

\subsection{Problem Formulation}
Our study aims to maximize the minimum data rate among all UEs in the network, ensuring fair access to resources. To achieve this, we define $C^\rho(\boldsymbol{p},\boldsymbol{e}) = \min (\boldsymbol{c_u}(t)),\,\, u \in N_\rho$, 
where $\boldsymbol{c_u}(t)$ 
is a vector of $c_{u,\rho}(t)$, and $\boldsymbol{p},\boldsymbol{e}$ indicates the vector of downlink transmission powers 
$p_{k,\rho}(t)$ and the matrix 
of RB allocation indicator $e_{u,k}$, respectively. 
We have thus formulated an optimization problem aiming for the best policy to jointly allocate resource blocks and transmission powers to UEs, constrained by the availability of network RBs and transmission powers to control interference.  
This methodology seeks to balance the diverse demands of UEs under variable channel conditions, optimizing the network's overall performance while adhering to operational constraints. 
\begin{subequations} 
\begin{align}\label{opt1}
 \argmax_{\boldsymbol{p},\boldsymbol{e}} & \hspace{0.5cm} 
  C^\rho(\boldsymbol{p},\boldsymbol{e}),\\
 \text{s.t.,} 
& \hspace{0.5cm}  \sum_{u=1}^{N_\rho}\sum_{k=1}^{K_\rho}e_{u,k} \leq K_\rho, \label{opt1_rb}\\
& \hspace{0.5cm}P_{min}\leq p_{k,\rho} \leq P_{max},\label{opt1_power}\\ 
& \hspace{0.5cm}  e_{u,k} \in \{0,1\}, \ \forall u \in N_\rho,\label{opt1_p}
\end{align}
\end{subequations}\vspace{-0cm}
where constraints \eqref{opt1_rb} and \eqref{opt1_power} define the limits on RB availability and transmission power, ensuring that interference management and allocations stay within the network's capacity. This underscores the challenge of allocating resources and power intelligently to ensure fair data rate distribution among UEs. Given the NP-hard nature and mixed-integer stochastic elements of this problem, direct solutions are challenging. Thus, applying the Markov decision process (MDP) framework to make informed decisions in uncertain conditions becomes vital. Translating the problem into an MDP and using dynamic resolution methods like DRL provides a strategic solution to this complex optimization issue.
\section{Proposed MAML-inspired Meta-RL framework}\label{MDRL}
To make a structured method for making informed decisions under uncertainty and using dynamic resolution techniques, we model the defined optimization problem \eqref{opt1} as an MDP. 

\subsection{MDP model}

The optimization problem \eqref{opt1} can be represented as an MDP with tuples $\langle \mathcal{S},\mathcal{A}, T,\gamma,r \rangle$, where $\mathcal{S}$, $\mathcal{A}$, and $T$ represent the state space, action space, and transition probability from the current state to the next state as $P(s_{t+1}|s_t)$, respectively, and $\gamma$ represents the discount factor and $r$ is reward function. The MDP tuples are described as follows:
\subsubsection{State}
At each timestep, $t$, the state $s_t \in \mathcal{S}$ represents the current O-RAN status, encompassing UEs' QoS metrics—average $Q_a$, minimum $Q_m$, and maximum $Q_x$
values, alongside the most recent resource $\boldsymbol{a}_{r,t-1}$ and power allocation actions $\boldsymbol{a}_{p,t-1}$. In our scenario, the UEs' QoS is defined as UEs' throughput. 
Hence, the agent's observation at time $t$ is denoted as $s_t = \{Q_a, Q_m, Q_x, \boldsymbol{a}_{r,t-1}, \boldsymbol{a}_{p,t-1}  \mid \forall u \in N_\rho\}$. 


\subsubsection{Action} 
At each timestep, $t$, the action vector $a_t \in \mathcal{A}$ specifies the required number of resource blocks and the necessary transmission power for each UE. Consequently, the agent employs its policy at time $t$ to select an optimal action, represented as $a_t = \{\boldsymbol{e}, \boldsymbol{p}\}$, where $\boldsymbol{e}$ denotes the indicator vector of RB's assigned to the UE and $\boldsymbol{p}$ represents the vector of transmission powers assigned to the UEs. 

\subsubsection{Reward} 
We incorporate reward and penalty components to design an efficient resource utilization function. This function assesses RB usage and transmission power to UEs, rewarding strategies that optimize resource and power efficiency while penalizing excess use to mitigate interference. Specifically, penalties are applied based on excess transmission power $\Tilde{p}_c = \sum_{u=1}^{N_\rho}\sum_{k=1}^{K_\rho} e_{u,k} p_{k,\rho}$ to prevent interference. To enforce resource conservation, a penalty is also imposed for exceeding the available RBs, calculated as $\Tilde{K}_r = \max(0,\sum_{u=1}^{N_\rho} \sum_{k=1}^{K_\rho} e_{u,k}-K_\rho) $. This method ensures the reward function promotes efficient allocation, enhances UE performance, and prevents excessive consumption of RBs. 
The reward $r_t$ combines \textit{sigmoid} functions of UEs' normalized minimum QoS, consumed power, and remaining RBs, is represented as $r_t = \text{sigmoid}(Q^n_m) - \text{sigmoid}(\Tilde{p}_c) - \text{sigmoid}(\Tilde{K}_r),$ 
with $Q^n_m = (Q_m - c_m)/(c_x-c_m)$, where $c_m$ and $c_x$ show the minimum and maximum values of service demands, representing normalized minimum QoS. This approach optimizes bandwidth and transmission power while meeting UEs' minimum QoS requirements. It guides the exploration of an MDP model using a DRL approach to discover the optimal policy $\pi^*(a_t|s_t;\theta_p)$ for RB allocation and power assignment.

Employing a DRL strategy, we address adaptive resource and power allocation in dynamic networks, optimizing performance within constraints and ensuring training stability. With the rise of virtualized base stations (vBS), we explore a meta-controller model in the RIC, enabling BS activation based on demand. This approach, directly informed by the O-RAN Alliance WG1's call for highly adaptable network management solutions as detailed in the~\cite{ORAN2023WhitePaper}, aims to create a scalable, flexible network adept at meeting changing demands. Unlike traditional DRL approaches, our meta-DRL framework incorporates rapid adaptability to new conditions, effectively addressing the shortcomings of existing DRL frameworks in managing the high variability and rapid dynamics typical of modern telecommunication networks. Further, we enrich our strategy by viewing each DU as an intelligent agent within a meta-learning framework, allowing for fast adaptation and learning from the network’s collective experiences. This approach is particularly beneficial in environments where network demands constantly evolve, necessitating the integration of new DUs to manage emerging hotspots or expanding service areas. The meta-DRL strategy not only responds to O-RAN WG1's emphasis on dynamically adjusting resource allocation in real time but also leverages the scalability of solutions to ensure continuous network performance improvement and resilience. \vspace{-0.cm}

\subsection{Proposed MAML-DRL approach}


We develop a novel framework that deploys distributed RL agents across the network to enhance stability and convergence, innovatively optimizing physical resource blocks and power allocation based on MAML-RL principles. Furthermore, to address the inaccuracies in power allocation identified in prior research~\cite{ji2023meta,yuan2021meta}, which arise from power quantization, we propose a continuous power allocation method using the Deep Deterministic Policy Gradient (DDPG) technique, aiming to enhance the accuracy and efficiency of system performance significantly.

In O-RAN wireless networks, which are equipped with RIC modules that provide sufficient computational resources, MAML outshines other meta-learning approaches like Reptile. It rapidly adapts to changes in the dynamic wireless environment through a few gradient steps, offering quicker and more efficient adjustments in complex RL scenarios~\cite{finn2017model}. This adaptability makes MAML particularly suitable for the fast-evolving conditions of wireless environments. In the context of meta-RL, a 'shot' refers to each distinct interaction or episode an agent uses to adapt to new tasks quickly. Few-shot learning involves the agent effectively handling new challenges with minimal prior exposure, typically from just a few interactions. This capability is critical for rapid adaptation in our network's continuously evolving demand scenarios. 
Further implementing our meta-learning strategy through a meta-controller within the near-real-time RIC module, functioning as an xApp, allows for continuous training of meta-learner RL agents across the network. When there is a need to integrate a new RL agent (i.e., a virtualized DU) due to increased UE density or critical conditions, the meta-controller facilitates the agent's adaptation, enabling more effective decision-making. \vspace{-0.1cm}
\begin{algorithm}[t!]
\SetAlgoLined
\textbf{Input}: $T$,\,\,$T_e$\,\,$N_g$,\,\,$\theta_{g},\forall g \in [0,N_g]$\,\,, $\theta_{M}$.  \\
\SetAlgoLined
\vspace{0.1cm}
\textbf{Meta training}\\
\For{iteration $t=1:T$}{
Initialize $\theta_{M} \to \theta_{g},\forall g \in [0,N_g]$.\\
\For{task $g=1:N_g$}{
\For{evaluation $e = 1 : T_e$}{
$r_g = \text{evaluate}(\pi_{p,g})$.\\
$\mathcal{B}_g\gets \langle s_t,a_t,s_{t+1},r_{g,t} \rangle $.
}
Update DDPG agent networks parameters $\theta_{t,g}$ using mini-batch experience as support set $\mathcal{B} ^{tr}_g$ based on \eqref{update_agent}.\\
Evaluate gradient of loss function of each task on mini-batch experience as query set $\mathcal{B} ^{val}_g$.
}
Update Meta-model network parameters $\theta_{t,M}$ using mini-batch experience of query set $\mathcal{B} ^{val}_g$ based on \eqref{update_meta}.
}\vspace{0.1cm}
\textbf{Meta adaptation}\\
Initialize $\theta_{M} \to \theta_{\text{new}}$.\\
\For{iteration $t=1:T_{\text{new}}$}{
$r_{\text{new}} = \text{evaluate}(\pi_{p,\text{new}})$.\\
$\mathcal{B}_\text{new}\gets \langle s_t,a_t,s_{t+1},r_{\text{new},t} \rangle $.\\
Update DDPG new agent network parameters $\theta_{t,\text{new}}$ using mini-batch experience as support set $\mathcal{B} ^{tr}_\text{new}$ based on \eqref{update_agent}.
}
\caption{The MAML-DRL algorithm }\vspace{-0.cm} 
\label{alg1}
\end{algorithm}\vspace{-0.2cm}
\setlength{\belowdisplayskip}{-10pt}
\setlength{\belowcaptionskip}{-3pt} 

\subsection{Meta-learning tasks}
Our goal in meta-learning is to train $N_g$ 
distinct meta-learner agents located in distributed DUs, each capable of quickly adapting to new tasks/environments with minimal training. These agents operate in unique environments, each with its own state spaces $\mathcal{S}$, action spaces $\mathcal{A}$, and customized reward functions $R_g$, preparing them for various environments and varying demands across network locations. By creating tasks that reflect the diversity of potential operational scenarios, we aim to improve our model's robustness and adaptability, allowing for swift adjustment to new challenges with minimal adaptation steps, as detailed in \cite{finn2017model}. $R_g = \{r_t | Q^n_m = \frac{Q_m - c_{m,g}}{c_{x,g}-c_{m,g}}\}, \forall g \in [0,N_g],$ shows each task specific reward function where $c_{m,g}$ are considered distinct for each individual environment.  
Our goal in meta-learning is to train $N_g$ 
Here, the meta-learner agent $g$ has a support set $\mathcal{B} ^{tr}_g$ to train the model within each task and a query set $\mathcal{B} ^{val}_g$ for weight updating of meta-model training. It helps to calculate the meta-objective, which is the loss computed on the query set predictions, and this meta-objective guides the update of the meta-parameters according to the following equations.  \vspace{-0.1cm}
\begin{align}
    \theta_{t,g} = \theta_{t-1,g} - \alpha \nabla_{\theta_g}\mathcal{L}(\theta_{t-1,g}, \beta^{tr}_g), \forall g \in [0,N_g], \label{update_agent}\\
    \theta_{t,M} = \theta_{t-1,M} - \alpha \nabla_{\theta_M}\sum_{g=1}^{N_g}\mathcal{L}(\theta_{t-1,g}, \beta^{val}_g), \label{update_meta}
\end{align}
where $\mathcal{L}(\theta_{t,g})$ indicates the loss function of DDPG at the agent $g$ in time $t$. Thus, \eqref{update_agent} updates the individual RL agent policy parameters based on the gradient descent method, and \eqref{update_meta} updates the meta-model policy parameters by aggregating the adaptation ability of the trained agents for each task. 


\subsection{Meta training}

Here, we organize the process into four main stages, focusing on the interactions and contributions of $N_g$ meta-learner agents, each employing a DDPG model within its environment.

\subsubsection{Initialization} Each meta-learner initializes its own DDPG model. Simultaneously, the meta-model is initialized separately to facilitate collective learning without direct interference.

\begin{table}[t!] 
	\footnotesize
	\centering
	\caption{\vspace*{-0cm} Simulation parameters}  \vspace{-0.2cm}
	\begin{tabular}{|>{\centering\arraybackslash}m{2.4cm}|>{\centering\arraybackslash}m{1.7cm}|>{\centering\arraybackslash}m{1.2cm}|>{\centering\arraybackslash}m{1.7cm}|}
		\hline
		\bf{Parameter} &\bf{Value } & \bf{Parameter} &\bf{Value }\\
		\hline
		Subcarrier spacing & $15$ kHz & $\sigma^2$ & $-173$ dBm \\
		\hline
		Total bandwidth/DU & \{$12$, $16$, $20$\}MHz  & $h$ & Rayleigh fading channel  \\
        \hline
		$K_\rho$ /DU & \{$60$, $80$, $100$\} & $T_e$ & $10$ \\
		\hline
		RB bandwidth/DU  & $200$ kHz & $N_g$ & $6$ \\
		\hline
		$P_{min}, P_{max}$  /RB & $ \{3$, $6$\} dBm & $T_{\text{new}}$ & $0.1 \times T$ \\
		\hline
		$N_\rho$ /DU & $30$ & batch size & $128$ \\
		\hline
	\end{tabular}\label{param} \vspace{-0.2cm}
\end{table}
\subsubsection{Training and Interaction} Meta-learners engage with their environments, gathering experiences (state, action, reward, next state) for learning. The learning process involves iterative updates to their actor and critic networks via DDPG, enhancing their decision-making capabilities based on observed outcomes.

\subsubsection{Contribution and Meta-update}

Upon completing a set number of episodes ($T_e$)
, meta-learners evaluate their advancements and relay their insights and effective gradients to the meta-model. This process ensures a collaborative update mechanism, enhancing the system's collective intelligence.

\subsubsection{Meta-model Optimization and Feedback} Integrating these contributions, the meta-model optimizes its learning strategy, enhancing adaptability and task performance. Enhanced strategies or parameters are then distributed to the meta-learners, supporting their continuous improvement and ability to tackle new challenges efficiently.


Algorithm \ref{alg1} outlines a MAML-inspired RL approach for addressing the joint RB and power allocation optimization problem, as defined in \eqref{opt1}-\eqref{opt1_p}. It describes a collaborative learning mechanism where individual learners (meta-learners) enrich a shared knowledge base (meta-model), improving each learner's efficiency and adaptability. \vspace{-0.1cm}

\begin{figure}[b]
\vspace{-10pt}
  \centering
    \includegraphics[width=0.71\columnwidth]{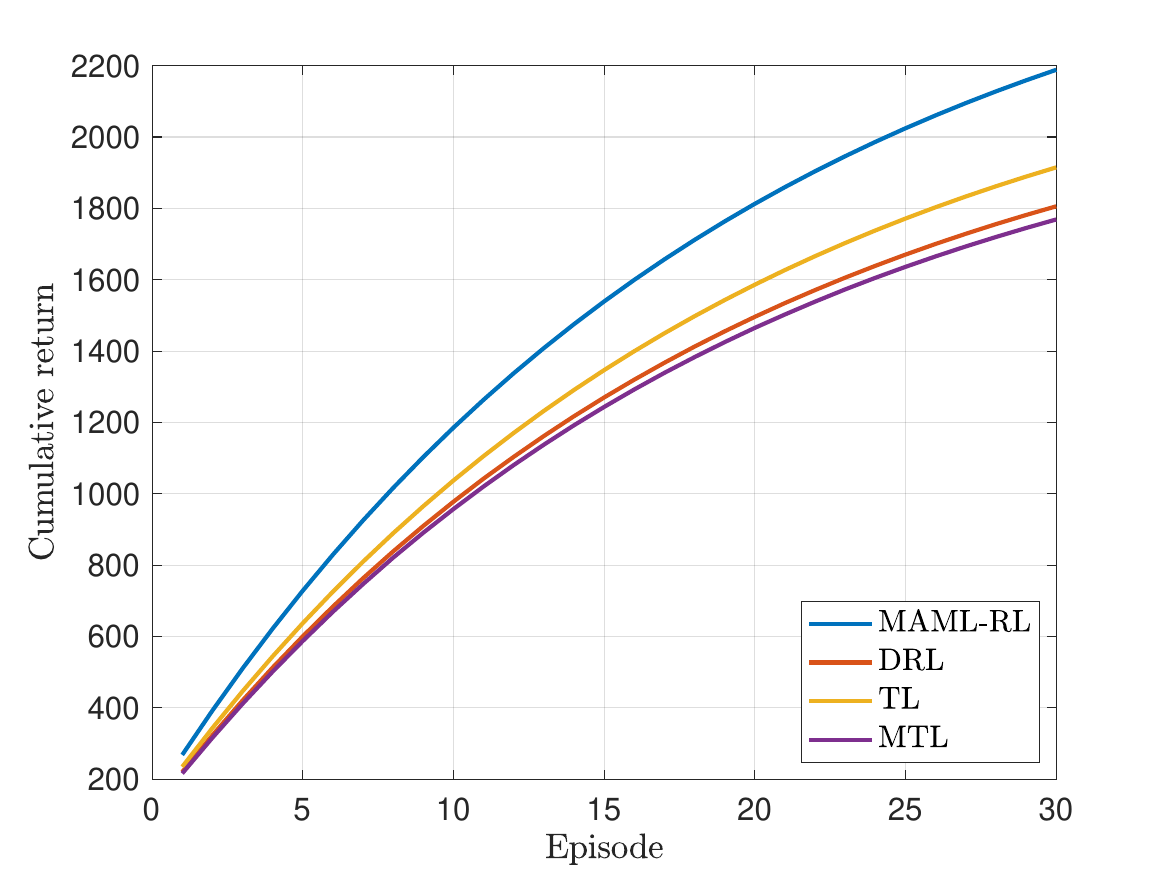}\vspace{-0.2cm}
    \caption{\small Performance comparison of the proposed MAML-RL and baseline algorithms. 
    }\vspace{-0.2cm}
    \label{Qos_performance}
\end{figure}
\begin{figure}[t!]
  \centering
    \includegraphics[width=0.71\columnwidth]{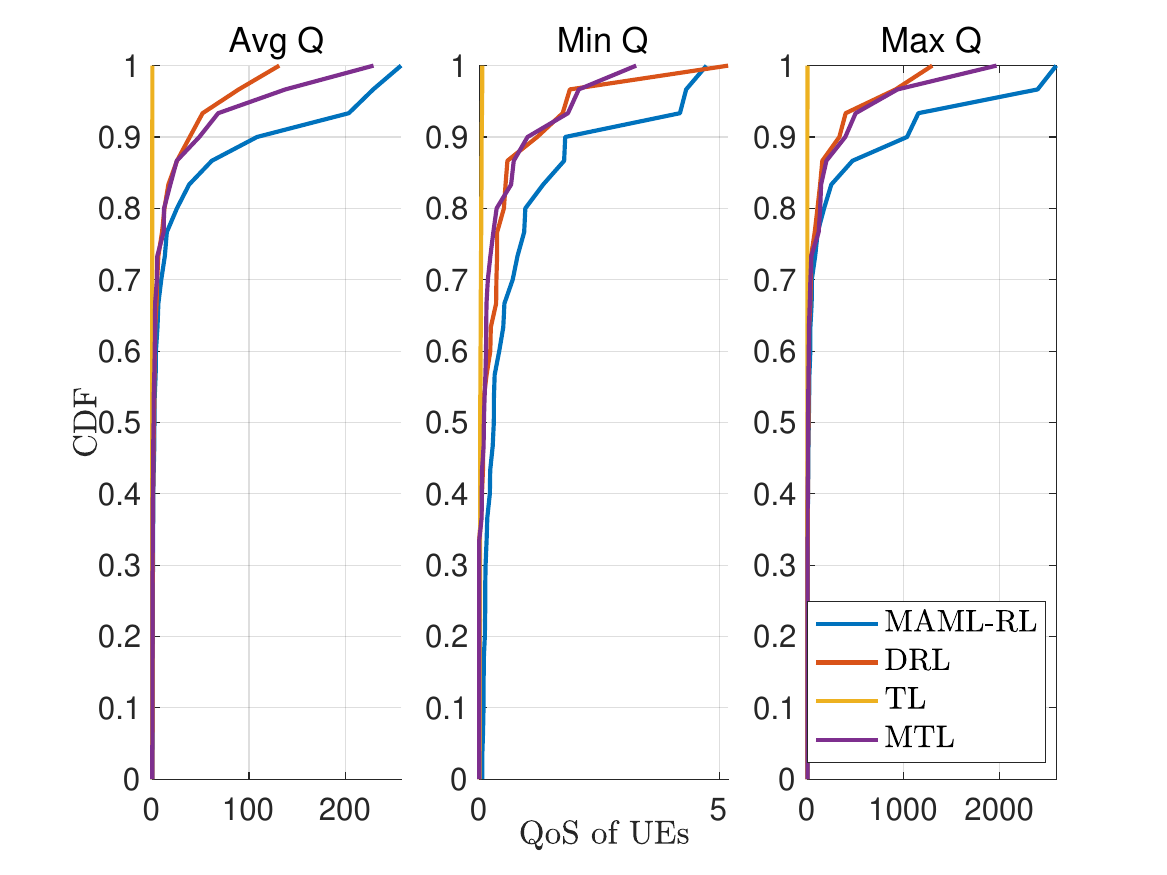}\vspace{-0.2cm}
    \caption{\small Performance comparison of the proposed MAML-RL approach on improving CDF of UEs' QoS values (avg, min and max) vs other baseline approaches. }\vspace{-0.2cm}
    \label{QoS_baseline}
\end{figure}
\section{Evaluation results}\label{sim}

In our O-RAN framework simulation for eMBB services, we model $N_g = 6$ distributed DUs covering different network areas, tasked with meta-learning varied tasks. The effectiveness of these tasks within the MAML-RL framework depends on their complexity and diversity. The framework caters to $30$ 
users per DU, uniformly and randomly spread across the network, serviced by dynamically allocated bandwidths of ${12, 16, 20}$ MHz or ${60, 80, 100}$ RBs. Users, exhibiting mobility with speeds between $10 m/s$ and $20 m/s$ and moving in one of seven possible directions ${\pm \pi/3, \pm \pi/6, \pm \pi/12, 0}$, traverse DU-assigned areas. Traffic profiles for these UEs shift among four levels—${idle, low, mid, high}$—with a $0.01$ transition probability each step, impacting RB allocation and bandwidth consumption based on real-time demands as outlined in~\cite{cheng2022reinforcement,lotfi2024open}. 
To deploy the DDPG strategy, we use a PyTorch-based actor-critic setup with three fully-connected layers of $300$, $400$, and $400$ neurons, utilizing \textit{tanh} functions and an \textit{Adam} optimizer at a $10^{-4}$ learning rate. Distributed as distinct task agents across the network, with a meta-model in the RIC module for information aggregation, our setup acknowledges non-uniform service demands, making DUs face varying traffic. We evaluate our MAML-RL method using a comprehensive set of benchmarks, starting DRL from scratch, transfer learning (TL) from a selected RL agent, and multi-task learning (MTL), which concurrently incorporates both a random task and a new task. This selection enables us to assess the flexibility and efficiency of our approach thoroughly. Spanning traditional to state-of-the-art techniques, TL and MTL are particularly noted for their advanced capabilities in RL. Detailed alongside other parameters in Table \ref{param}, these benchmarks facilitate a robust comparison of learning strategies within our framework. 

Fig. \ref{Qos_performance} compares the cumulative return achieved by various RL approaches across several episodes. The results were measured with $\gamma=0.99$ and averaged over a sufficient number of runs. As the graph shows, the proposed MAML-RL approach can provide up to $19.8\%$ gain over other baseline approaches that show the efficiency of the proposed approach. 

\begin{figure}[t!]
  \centering
    \includegraphics[width=0.71\columnwidth]{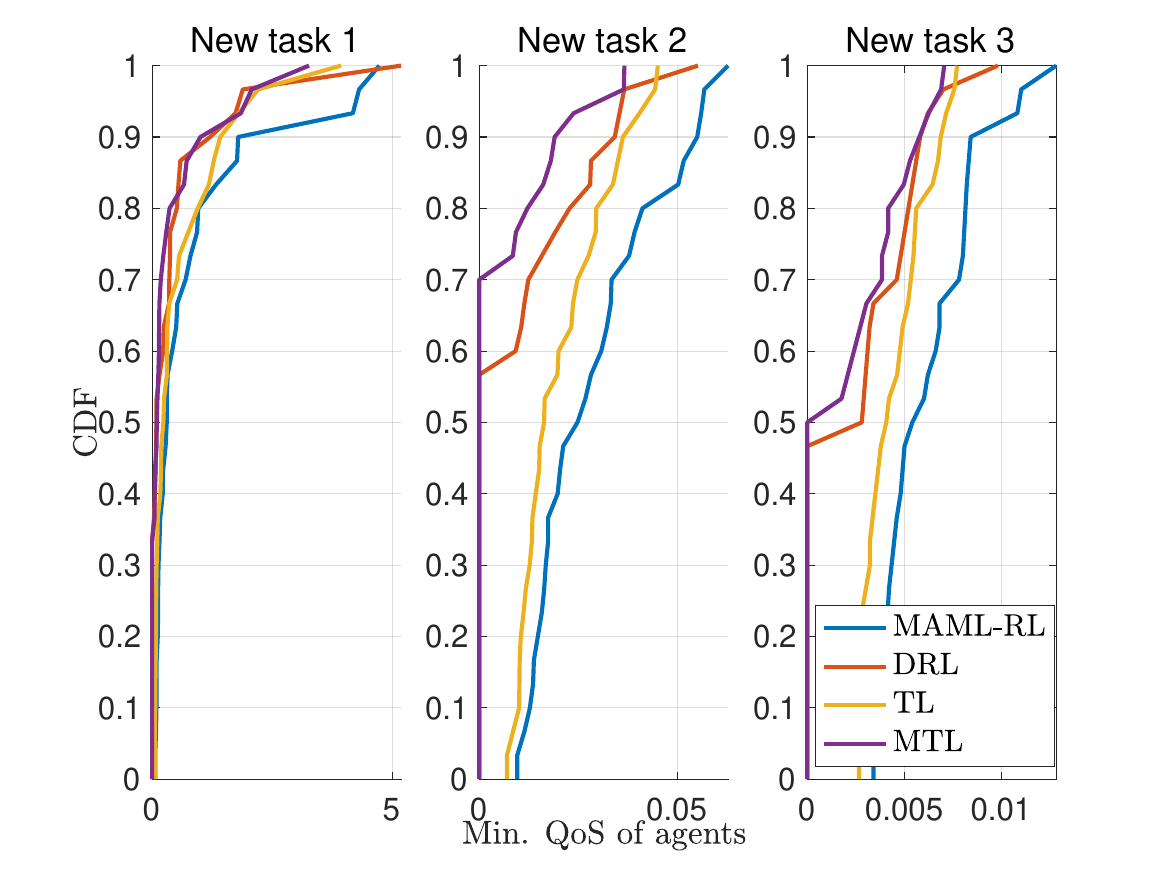}\vspace{-0.2cm}
    \caption{\small Performance comparison of CDF of the minimum QoS of all UEs in three distinct new tasks vs DRL training from scratch. 
    }\vspace{-0.2cm}
    \label{QoS_tasks}
\end{figure}
\begin{figure}[t!]
  \centering
    \includegraphics[width=0.71\columnwidth]{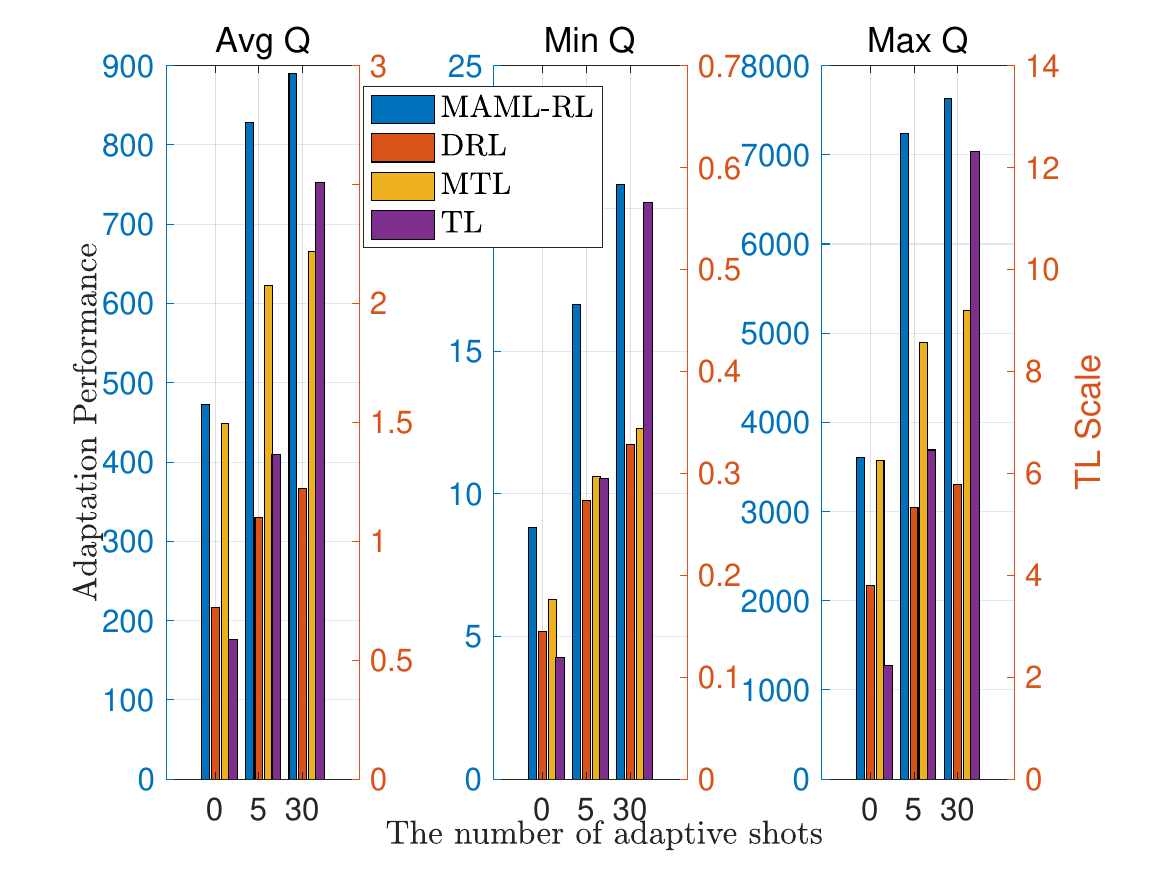}\vspace{-0.2cm}
    \caption{\small Adaptation performance for average, minimum and maximum values of UEs' QoS for the different number of adaptive shots. 
    }\vspace{-0.3cm}
    \label{adapt_performance}
\end{figure}

The cumulative distribution function (CDF) graphs in Fig. \ref{QoS_baseline} contrast the throughput of UEs using MAML-RL, DRL, TL, and MTL. The proposed MAML-RL outperforms these models in average, minimum, and maximum throughput, proving its robustness and reliability for UE throughput. While DRL and MTL are robust in maximum throughput, suggesting they can reach MAML-RL levels sometimes, TL provides lower throughput, hinting at its unsuitability for heterogeneous UE scenarios.

Fig. \ref{QoS_tasks} shows the CDF for the minimum QoS of UEs across three new task scenarios, each with unique UE starting conditions and available RBs. These scenarios test RL approaches against the complex backdrop of dynamic wireless channels, traffic, and UE mobility. MAML-RL is the most robust and efficient, leading in nearly all tasks. DRL starts well but falls behind as QoS demands rise. TL is competent but not outstanding, while MTL trails, showing less adaptability to task diversity. This variety in tasks highlights each method's inherent efficiency and adaptability to new challenges. 

Fig. \ref{adapt_performance} presents the adaptation performance for MAML-RL, DRL, TL, and MTL against average, minimum, and maximum QoS metrics in different adaptive shots. MAML-RL stands out, showing superior performance, with more noticeable gains as adaptive shots increase. DRL and MTL are competitive, yet neither matches MAML-RL's level. Conversely, TL consistently shows the lowest adaptability across the metrics. 

\begin{figure}[t!]
  \centering
    \includegraphics[width=0.7\columnwidth]{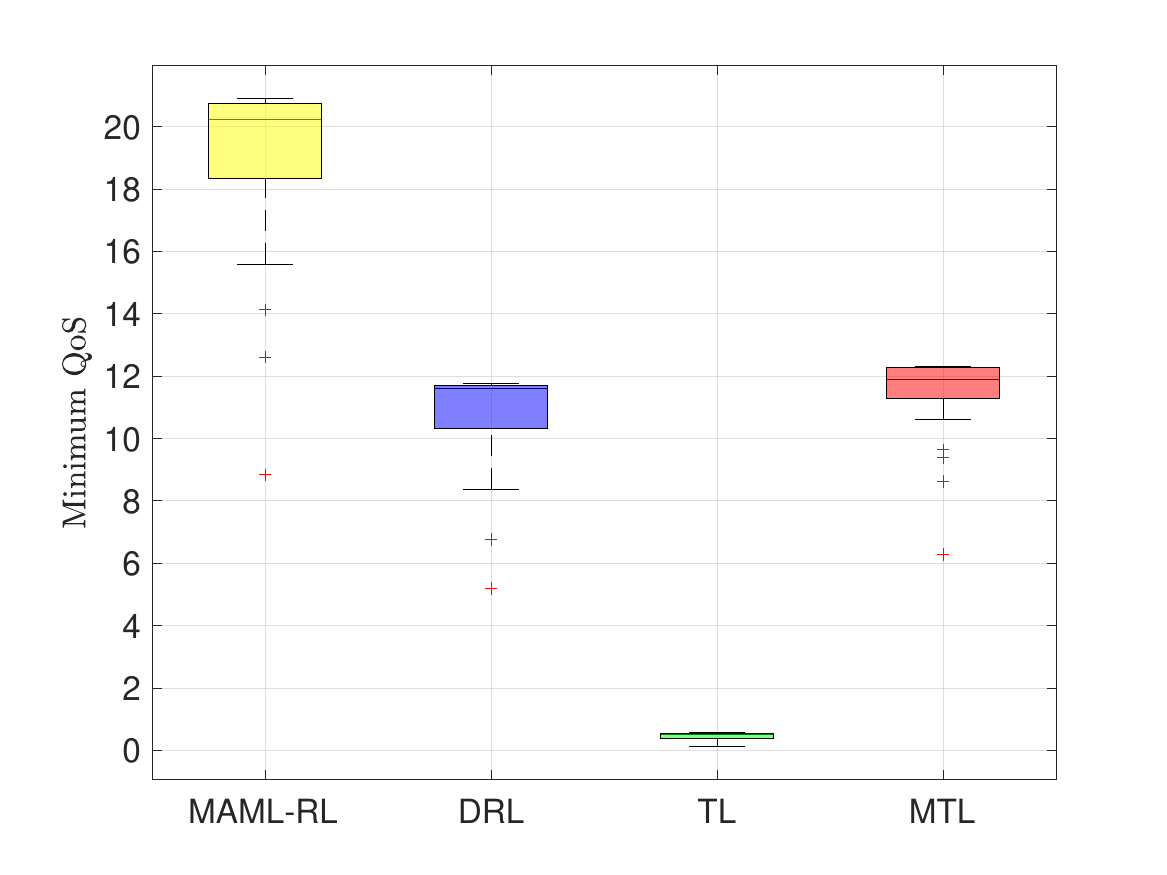}\vspace{-0.2cm}
    \caption{\small Distribution of minimum QoS of UEs across the network in 5 shots scenario using different methods. }\vspace{-0.2cm}
    \label{boxplot}
\end{figure}

Fig. \ref{boxplot} shows how MAML-RL surpasses other methods in achieving the maximum value of minimum QoS, with its distribution's upper range and outliers indicating higher QoS instances. DRL's results are less varied and have a lower median QoS than MAML-RL. TL exhibits limited variability with lower QoS, while MTL, despite being the weakest overall, shows sporadic peaks in QoS. The graph highlights MAML-RL's ability to attain superior QoS levels and exceptional performance potential.\vspace{-0.1cm}



\section{Conclusion}\label{conclusion}
In the emerging O-RAN technology, optimizing resource and power allocation presents significant challenges due to the dynamic nature of wireless networks. Addressing this, we introduce a novel Meta-DRL approach inspired by MAML, which primarily enhances the adaptivity to dynamic network conditions. Our approach achieves a substantial $19.8\%$ improvement over conventional methods, 
affirming the capability of the proposed architecture to increase the efficiency of wireless networks and facilitate quick adaptation to dynamic conditions. This research enhances network optimization by improving generalization and rapid adaptivity for new decision-making agents within the O-RAN ecosystem, which is crucial for advanced network management. It also opens avenues for applications such as autonomous network repairs and dynamic spectrum management, illustrating its potential impact on adaptive technologies in complex environments.


\vspace{-0.1cm}

\def\baselinestretch{0.83}
\bibliographystyle{IEEEbib}
\bibliography{Main}
\end{document}